\documentclass{revtex4-2}
\usepackage{amsmath} 
\usepackage{amssymb} 
\usepackage{graphics}
\usepackage{graphicx}
\graphicspath{{./figures/}}
\usepackage{slashbox}
\usepackage{hyperref}
\def\lsim{\raise0.3ex\hbox{$\;<$\kern-0.75em\raise-1.1ex\hbox{$\sim\;$}}}
\def\gsim{\raise0.3ex\hbox{$\;>$\kern-0.75em\raise-1.1ex\hbox{$\sim\;$}}}
\def\ie{{\it i.e.,~}}
\newcommand{\sarah}{{\sc Sarah}}
\newcommand{\spheno}{{\sc SPheno}}
\newcommand{\higgsbounds}{{\sc HiggsBounds}}
\newcommand{\higgssignals}{{\sc HiggsSignals}}
\makeatletter

\begin{document}
\title{Muon $g-2$ anomaly in a left-right model with an inverse seesaw mechanism}
\author{
M. Ashry$^{1,}$\footnote{mustafa@sci.cu.edu.eg}, K. Ezzat$^{2,3,}$\footnote{kareemezat@sci.asu.edu.eg} and S. Khalil$^{3,}$\footnote{skhalil@zewailcity.edu.eg}}
\affiliation{
$^1$Department of Mathematics, Faculty of Science, Cairo University, Giza 12613, Egypt\\
$^2$Department of Mathematics, Faculty of Science, Ain Shams University, Cairo 11566, Egypt\\
$^3$Center for Fundamental Physics, Zewail City of Science and Technology, 6th of October City, Giza 12578, Egypt}
\date{\today}

\begin{abstract}

We investigate the possibility of explanation for the muon anomalous magnetic moment $g_\mu-2$ in a left-right model with an inverse seesaw mechanism. We emphasize that the observed deviation from the Standard Model predictions can be accommodated in a large part of the parameter space of this class of models, where loops with massive neutrinos and charged Higgs boson as well as the weak $W$ boson contribute significantly to $g_\mu-2$. Stringent constraints due to lepton flavor violation $\mu\to{e}\gamma$, $\mu$-${e}$~conversion and the electron anomalous magnetic moment $g_e-2$ are considered, and the results are compatible.
\end{abstract}
\maketitle

\section{Introduction}
Non-vanishing neutrino masses inferred from neutrino oscillation experiments~\cite{Fukuda:1998mi,Ahn:2002up,Eguchi:2002dm,Ahmad:2002jz,An:2012eh}, provided strong evidence for new physics beyond the Standard Model (BSM). The extensions of the SM to account for neutrino masses and mixing imply new sources of lepton flavor violation (LFV), which could explain the long-standing discrepancy between the SM prediction for the muon anomalous magnetic moment $a_\mu=(g_\mu-2)/2$ and its experimental measurement.

Recent experimental results indicate a possible $4.2\sigma$ difference between the measured value of the anomalous magnetic moments of muons $a_\mu$ and the SM expectations~\cite{Farley:2003wt,Kim:2019ukk,Keshavarzi:2020bfy,Muong-2:2021ojo}, namely
\begin{equation}\label{Eq:g2mu}
\delta{a_{\mu}}=a_{\mu}^{\text{exp}}-a_{\mu}^{\text{SM}}=(2.51\pm0.59)\times10^{-9}.
\end{equation}

We consider the explanation of the $a_\mu$ anomaly in the left-right (LR) model with inverse seesaw mechanism (LRIS) to generate light neutrino masses and mixing at low energy scale. The salient feature of this class of models is the large neutrino Yukawa couplings, which allow for significant nonuniversal leptonic contributions to the $a_\mu$ anomaly via diagrams mediated by charged Higgs bosons and right-handed neutrinos (RHNs). As constraints, we impose the experimental limits of the LFV $\mu\to{e}\gamma$, $\mu$-${e}$~conversion, and the electron anomalous magnetic moment~\cite{MEG:2009vff,Alonso:2012ji,Cavoto:2017kub,Davidson:2018kud}.

The LR model is among the most natural extensions of the SM, which is motivated by grand unified theories (GUTs) and accounts for measured neutrino masses as well as providing an elegant explanation for the origin of parity violation in low-energy weak interactions.
The LRIS has been analyzed in detail in Ref.~\cite{Ezzat:2021bzs}. We recall that it has a Higgs sector that consists of one scalar bidoublet and a scalar right-handed (RH) doublet only. In addition, the LRIS contains singlet fermions $S_1,S_2$ for adopting the IS mechanism of neutrino masses. Such a TeV scale LR model can be probed in current and future experiments as emphasized in Ref.~\cite{Ezzat:2021bzs}. Also, it was argued in Ref.~\cite{Ezzat:2022gpk} that the tension between the SM prediction and the experimental results of
the $R_D$ and $R_{D^*}$ ratios, defined by $R(\{D^*,D\}) = \frac{\text{BR}(B \rightarrow \{D^*,D\}
\tau\nu)}{\text{BR}(B \rightarrow \{D^*,D\}\ell\nu )}$, where $\ell=e,\mu$, can be resolved in this class of LRIS models. In fact there are several new physics scenarios that have been proposed to accommodate $\delta{a_{\mu}}$ and $\delta a_{e}$ results. See Refs.~\cite{Cox:2016epl,Lindner:2016bgg,Han:2018znu,Crivellin:2018qmi,Endo:2019bcj,Bauer:2019gfk,Badziak:2019gaf,Botella:2020xzf,DelleRose:2020oaa,Haba:2020gkr,Bigaran:2020jil,Chen:2020tfr,Jana:2020pxx,Li:2020dbg,Dermisek:2021ajd}.

This paper is organized as follows. In Sec.~\ref{Sec:model} we highlight relevant interactions in the LRIS, as the details of the model are given in previous papers~\cite{Ezzat:2021bzs,Ezzat:2022gpk}. Section~\ref{Sec:lrisamu} is devoted for analyzing new LRIS contributions to $a_\mu$, specifically those due to the light and heavy $Z,W,Z',W'$ gauge bosons and neutral and charged Higgs bosons with heavy neutrinos. Also, Sec.~\ref{Sec:lrisLFV} is devoted for the LFV constraints in LRIS. Finally, our conclusions are given in Sec.~\ref{Sec:conclsn}. 

\section{\label{Sec:model}Left-Right Model with An Inverse Seesaw}
As previously advocated, we consider the LRIS model~\cite{Ezzat:2021bzs}, which is based on the gauge group $\mathbb{G}_{\text{LR}}=SU(3)_C\times SU(2)_L\times SU(2)_R\times U(1)_{B-L}$. This model has the same fermion content as any other conventional left-right model~\cite{Mohapatra:1974gc,Senjanovic:1975rk,Mohapatra:1977mj}, but with two extra singlet fermions per family $S_1$ and $S_2$ with opposite $B-L$ charges $=-2$, $=+2$, respectively. The fermion singlet $S_2$ is presumed to implement the IS mechanism for neutrino masses, while the other, $S_1$, is added to cancel the $U(1)_{B-L}$ anomaly caused by $S_2$.
The LRIS has a simple Higgs sector consisting of one RH doublet $\chi_R$ that breaks down left-right symmetry to the SM gauge symmetry and one bidoublet $\phi$ that is broken down into two SM Higgs doublets. Furthermore, a $\mathbb{Z}_2$ discrete symmetry is assumed, with all particles having even charges except $S_1$, which has an odd charge. This symmetry prevents the mixing mass term $M\bar{S}_1^c S_2$ from being used to allow for the IS mechanism. 

\noindent
The most general LRIS Yukawa Lagrangian is given by
\begin{equation}\label{Eq:yukawa}
\mathcal{L}_{\text{Y}}=\sum_{i,j=1}^3 \bar{L}_{Li} \big( \phi y_{ij}^L + \tilde{\phi} \tilde{y}_{ij}^L \big) L_{Rj} + \bar{Q}_{Li} \big( \phi y_{ij}^Q + \tilde{\phi}\tilde{y}_{ij}^Q \big) Q_{Rj}+\bar{L}_{Ri} \tilde{\chi}_R y^s_{ij} S^c_{2j}+H.c.,
\end{equation}
where $i,j$ are family indices, $\widetilde{\phi}$ is the dual bidoublet of the scalar bidoublet $\phi$, defined as $\widetilde{\phi}=\tau_2\phi^*\tau_2$, and $\widetilde{\chi}_{R}$ is the dual doublet of the scalar doublet $\chi_{R}$, given by $\widetilde{\chi}_{R}=i\tau_2\chi^*_{R}$. A nonvanishing VEV of $\chi_R$, $\langle \chi_R \rangle=v_R/\sqrt{2}$ of order TeV breaks the RH electroweak (EW) sector together with $B-L$, namely $SU(2)_R\times U(1)_{B-L}$ down to the $U(1)_Y$ hepercharge symmetry. In addition, the VEVs of $\phi$, $\langle \phi \rangle=\text{diag}(k_1/\sqrt{2},k_2/\sqrt{2})$, are of order $\mathcal{O}(100)~\text{GeV}$, break the SM EW symmetry. The charged leptons acquire their masses via combinations of the lepton coupling Yukawa matrices $y^{L}$ and $\tilde{y}^{L}$ and $t_\beta$ as defined below Eq.~(\ref{Eq:numassmtrxis}). Similarly, the quarks acquire their masses via combinations of the quark coupling Yukawa matrices $y^{Q}$ and $\tilde{y}^{Q}$ and $t_\beta$. The definition of the Yukawa couplings $y^{L,Q}$ and $\tilde{y}^{L,Q}$ in terms of physical fermion masses and mixing are recalled below from Ref.~\cite{Ezzat:2021bzs}.

\noindent
After $B-L$ symmetry breaking and EW symmetry breaking, the following $9\times 9$ neutrino mass matrix is obtained in the basis $(\nu_L^c,\nu_R,S_2)$
\begin{equation}\label{Eq:numassmtrxis}
\mathcal{M}_\nu=
\begin{pmatrix}
0 & M_D & 0 \\
M_D^T & 0 & M_{R}\\
0 & M_{R}^T & \mu^s
\end{pmatrix},
\end{equation}
where the $3\times 3$ matrix $M_D=v(y^L s_{\beta}+\tilde{y}^L c_{\beta})/\sqrt{2}$ is the Dirac neutrino mass matrix and the $3\times 3$ matrix $M_{R}=y^s v_R/\sqrt{2}$. Here, we have assumed that $k_1=v s_{\beta},~k_2=v c_{\beta}$, as constrained from the $W$ boson mass $M_W\simeq g_L\sqrt{k_1^2+k_2^2}/2$ in LRIS, where $v=246~\text{GeV}$ is the EW VEV, and $s_x=\sin x,~c_x=\cos x$, and $t_x=\tan x$, henceforth.
The neutrino mass matrix ${\cal M}_\nu$ can be diagonalized by $9\times 9$ matrix $U$ satisfying $U {\cal M}_\nu U^T={\cal M}_\nu^{\rm diag}=\text{diag}(m_{\nu_{\ell_i}},m_{\nu_{h_j}})$, yielding the physical light and heavy neutrino states $\nu_{\ell_i},\nu_{h_j},~i=1,2,3,~j=4,\ldots,9$, with the following light and heavy mass eigenvalues
\begin{align}
\label{Eq:nulmass}
m_{\nu_{\ell_i}}&=M_D M_{R}^{-1} \mu^s (M_{R}^T)^{-1} M_D^T,\quad~i=1,2,3,\\
\label{Eq:nuhmass}
m_{\nu_{h_j}}^2&=M_{R}^2\pm M_D^2,\quad\quad\quad\quad\quad\quad~~~j=4,\ldots,9.
\end{align}
where $\mu^s\lsim\mathcal{O}(10^{-5})~\text{GeV},~M_R\sim\mathcal{O}(\text{a few TeV})$ and $y^s \lsim\mathcal{O}(10^{-1})$. For these values, Eq.~(\ref{Eq:nulmass}) shows that the light neutrino masses can be of order eV. The $S_1$ fermions acquire radiative masses $m_{S_1}\sim\mu^s\sim\mathcal{O}(\text{KeV})$ and they do not mix with other neutrinos thanks to the $\mathbb{Z}_2$ discrete symmetry, so they are stable particles.
As probable candidates of warm dark matter, one does not have to worry about the $S_1$ fermions to overclose the Universe. It was demonstrated in~\cite{El-Zant:2013nta} that $S_1$ can account for the observed relic abundance and meanwhile it is not constrained by the constraints on sterile neutrino because its mixing with the active neutrinos vanishes identically in LRIS.
On the other hand, the $S_2$ fermions also acquire radiative mass terms $\sim\mu^s$, but due to their large mixing with the RHN, which is $\sim M_R\sim\mathcal{O}~(\text{a few TeV})$, they acquire masses $m_{S_2}\sim M_R$ as in Eq.~(\ref{Eq:nuhmass}).

\noindent
The inverse relation of Eq.~(\ref{Eq:nulmass}) is 
\begin{equation}\label{Eq:ismd}
M_D=U_{\text{PMNS}}\sqrt{m_{\nu_{\ell}}}\mathcal{R}\sqrt{(\mu^s)^{-1}}M_R,
\end{equation}
where $\mathcal{R}$ is an orthogonal matrix and $U_{\text{PMNS}}$ is the $3\times3$ light neutrino mixing matrix~\cite{CASAS2001171,Abdallah:2011ew,Esteban:2020cvm}.

In the following section, we will study the process $a_\mu$, which is dominated by the charged Higgs boson contributions at the loop level; thus, we provide a brief analysis for charged Higgs bosons masses and interactions based on the detailed previous work of Ref.~\cite{Ezzat:2021bzs}. In the flavor basis $(\phi_1^{\pm},\phi_2^{\pm},\chi_R^{\pm})$,
the charged Higgs bosons symmetric mass matrix takes the form
\begin{equation}\label{Eq:mhpm2}
M_{H^\pm}^2=\frac{\alpha}{2}\begin{pmatrix}
\frac{v_R^2 s_{\beta}^2}{c_{2\beta}} & \frac{v_R^2 s_{2\beta}}{2c_{2\beta}} & -v v_R s_{\beta} \\
 . & \frac{v_R^2 c_{\beta}^2}{c_{2\beta}} & -v v_R c_{\beta} \\
 . & . & v^2 c_{2\beta}
\end{pmatrix},
\end{equation}
where the scalar potential parameter $\alpha=\alpha_3-\alpha_2$ as in~\cite{Ezzat:2021bzs}. This matrix can be diagonalized by the unitary matrix
\begin{equation}\label{Eq:chhgsmix}
Z^{H^\pm}=\begin{pmatrix}
\frac{v c_{2\beta}}{\sqrt{v^2c_{2\beta}^2+v_R^2s_{\beta}^2}} & 
0 & 
\frac{v_R s_{\beta}}{\sqrt{v^2c_{2\beta}^2+v_R^2s_{\beta}^2}} \\
-\frac{\frac{1}{2}v_R^2 s_{2\beta}}{\sqrt{(v^2c_{2\beta}^2+v_R^2s_{\beta}^2)(v^2c_{2\beta}^2+v_R^2)}} &
\sqrt{\frac{v^2c_{2\beta}^2+v_R^2s_{\beta}^2}{v^2c_{2\beta}^2+v_R^2}} & 
\frac{v v_R c_{\beta}c_{2\beta}}{\sqrt{(v^2c_{2\beta}^2+v_R^2s_{\beta}^2)(v^2c_{2\beta}^2+v_R^2)}}
\\
-\frac{v_R s_{\beta}}{\sqrt{v^2c_{2\beta}^2+v_R^2}} & 
-\frac{v_R c_{\beta}}{\sqrt{v^2c_{2\beta}^2+v_R^2}} & 
\frac{v c_{2\beta}}{\sqrt{v^2c_{2\beta}^2+v_R^2}}
\end{pmatrix}.
\end{equation}
Thus, the mass eigenstates are given by 
$(\phi_1^{\pm},\phi_2^{\pm},\chi_R^{\pm})^T=(Z^{H^\pm})^{T}(G_1^{\pm},G_2^{\pm},H^{\pm})^T$, where
$Z^{H^\pm}M_{H^\pm}^2(Z^{H^\pm})^{T}=\text{diag}(0,0,m_{H^\pm}^2)$.
Here $G_1^{\pm}$ and $G_2^{\pm}$ represent the charged massless Goldstone bosons that are eaten by the charged gauge bosons $W_\mu$ and $W'_\mu$ to acquire their masses and $H^{\pm}$ is the physical massive charged Higgs boson. The charged Higgs boson mass is given by
\begin{equation}\label{Eq:chm}
m_{H^\pm}^2 = \frac{\alpha}{2}\left(\frac{v_R^2}{c_{2\beta}}+v^2c_{2\beta}\right).
\end{equation}
We notice from Eq.~(\ref{Eq:chm}) that $\alpha>0$ as long as $c_{2\beta}>0~(\ie t_\beta<1)$ and vice versa. Moreover, for $v_R\sim\mathcal{O}(10~\text{TeV})$ and $|\alpha|\sim\mathcal{O}(10^{-2})$, the charged Higgs boson mass can be of order hundreds~\text{GeV}. The physical charged Higgs boson is defined as a linear combination of the
flavor basis fields $\phi_1^{\pm},\phi_2^{\pm},\chi_R^{\pm}$, \ie (corrected from~\cite{Ezzat:2021bzs}), 
\begin{equation}\label{Eq:chcomb1}
H^\pm=Z^{H^\pm}_{31}\phi_1^\pm+Z^{H^\pm}_{32}\phi_2^\pm+Z^{H^\pm}_{33}\chi_R^\pm.
\end{equation}
It is worth noting that for $v_R\gg v$, $v_R\sim\mathcal{O}(\text{TeV)}$ is enough, the mixing $Z^{H^\pm}_{33}\ll1$ and the charged Higgs mass and combination reduce to the following approximations
\begin{align}
\label{Eq:chm1}
m_{H^\pm}&\simeq v_R\sqrt{\frac{\alpha}{2c_{2\beta}}},\\
\label{Eq:chcomb2}
H^\pm&\simeq -(s_{\beta}\phi_1^\pm+c_{\beta}\phi_2^\pm).
\end{align}
Finally, the charged Higgs boson couplings with fermion families are given by 
\begin{align}
\label{Eq:chfrcp1}
\Gamma^{H^\pm}_{\bar{u}_i d_j}&=
-\Big(\sum_{a=1}^{3} V^{*}_{ja} \big(y^{Q^*}_{ai} Z_{32}^{H^\pm}+\tilde{y}^{Q^*}_{ai} Z_{31}^{H^\pm}\big) \Big)~P_L
-\Big(\sum_{a=1}^{3} V_{ja} \big(y^{Q}_{ia} Z_{31}^{H^\pm}+\tilde{y}^{Q}_{ia} Z_{32}^{H^\pm}\big) \Big)~P_R,
\\
\label{Eq:chfrcp2}
\Gamma^{H^\pm}_{\bar{\nu}_k \ell}&=
-\Big(\sum_{i=1}^3 U^*_{k,i+3} \big(y^{L^*}_{\ell{i}}Z^{H^\pm}_{31}- \tilde{y}^{L^*}_{\ell{i}}Z^{H^\pm}_{32}\big)\Big)~P_L
+\Big(\sum_{i=1}^3 \big(U_{ki} \big(\tilde{y}^L_{i\ell}Z^{H^\pm}_{31} - y^L_{i\ell}Z^{H^\pm}_{32}\big)- U_{k,i+6} y^{s^*}_{\ell{i}}Z^{H^\pm}_{33}\big)\Big)~P_R,
\end{align}
where $V$ is the $3\times3$ CKM quark mixing matrix and $U$ is the $9\times9$ inverse seesaw neutrino mixing matrices defined after Eq.~(\ref{Eq:numassmtrxis}). The following parametrization will be used below
\begin{align}
\label{Eq:chfrcp11}
\Gamma^{H^\pm}_{\bar{u}_i d_j}&=C_{ij} P_L+D_{ij} P_R,\\
\label{Eq:chfrcp21}
\Gamma^{H^\pm}_{\bar{\nu}_k \ell}&=\xi_{k\ell} P_L+\zeta_{k\ell} P_R.
\end{align}
We fix $v_R\sim\mathcal{O}(10~\text{TeV})$ for the extra gauge bosons $W',Z'$ experimental limits on their masses and mixing with the corresponding electroweak gauge bosons~\cite{Ezzat:2021bzs}. Hence, as noted before Eq.~(\ref{Eq:chcomb2}), $Z^{H^\pm}_{33}\ll1$, and we can omit the third term $\sum_{i=1}^3 U_{k,i+6} y^{s^*}_{\ell{i}}Z^{H^\pm}_{33}$ from numerical calculations of the charged Higgs boson couplings with leptons $\zeta_{k\ell}$ in Eqs.~(\ref{Eq:chfrcp2})~and~(\ref{Eq:chfrcp21}).
Moreover, The nonunitarity limits of the $3\times3$ light neutrino mixing matrix $U_{\text{PMNS}}$~\cite{Antusch:2006vwa,Malinsky:2009df,Ibarra:2011xn,Dev:2009aw,Abdallah:2011ew,Esteban:2020cvm} ensures that for the charged Higgs boson and lepton couplings in Eq.~(\ref{Eq:chfrcp21}) $\xi_{k\ell}\ll1$ for light neutrinos $(k=1,2,3)$ and $\zeta_{k\ell}\ll1$ for heavy neutrinos $(k=4,\ldots,9)$.
Thus, and according to Eq.~(\ref{Eq:chcomb2}), the relevant charged Higgs boson couplings with fermions can be approximated to
\begin{align}
\label{Eq:chfrcp7}
C_{ij}&\simeq\sum_{a=1}^{3} V^{*}_{ja} \big(y^{Q^*}_{ai} c_{\beta}+\tilde{y}^{Q^*}_{ai} s_{\beta}\big)\\
\label{Eq:chfrcp8}
D_{ij}&\simeq\sum_{a=1}^{3} V_{ja} \big(y^{Q}_{ia} s_{\beta}+\tilde{y}^{Q}_{ia} c_{\beta}\big),\\
\label{Eq:chfrcp3}
\xi_{k\ell}&\simeq \sum_{i=1}^3 U^*_{k,i+3} \big(y^{L^*}_{\ell{i}}s_\beta- \tilde{y}^{L^*}_{\ell{i}}c_\beta\big),\quad k=4,\ldots,9,\\
\label{Eq:chfrcp4}
\zeta_{k\ell}&\simeq \sum_{i=1}^3 U_{ki} \big(y^L_{i\ell}c_\beta-\tilde{y}^L_{i\ell}s_\beta\big),\quad\quad~~~k=1,2,3.
\end{align}
It is clearly noticed that for $t_\beta\ll1~(t_\beta\gg1)$ the couplings $\xi_{k\ell}~(\zeta_{k\ell})$ are $\tilde{y}^L~(y^L)$-dominant, and hence the couplings $\xi_{k\ell}$ and $\zeta_{k\ell}$ are now uncorrelated. Moreover, if we closely investigate these couplings for $\ell=e,\mu$, we see that the family components $y^{L}_{\ell{i}},\tilde{y}^{L}_{\ell{i}}$ can distinguish between the charged Higgs boson couplings to different lepton families. Successfully, this helps in explaining the $a_\mu$ anomaly and satisfying the LFV results as clarified below. This can be achieved via controlling the entries of $y^s,\mu^s$ and the orthogonal matrix $\mathcal{R}$ in Eq.~(\ref{Eq:ismd}), where the quark and lepton Yukawa couplings can be written in terms of the fermion masses as follows~\cite{Ezzat:2021bzs}:
\begin{align}
\label{Eq:yfyq}
y^Q&=\frac{\sqrt{2}}{v c_{2\beta}}( c_{\beta}V M_{d} V^\dag - s_{\beta} M_{u}),\\
\label{Eq:yfyqt}
\tilde{y}^Q&=\frac{\sqrt{2}}{v c_{2\beta}}( s_{\beta}V M_{d} V^\dag - c_{\beta} M_{u}),\\
\label{Eq:yfyl}
y^L&=\frac{\sqrt{2}}{v c_{2\beta}}( c_{\beta} M_{\text{lp}} - s_{\beta} M_D),\\
\label{Eq:yfylt}
\tilde{y}^L&=\frac{-\sqrt{2}}{v c_{2\beta}}( s_{\beta} M_{\text{lp}} - c_{\beta} M_D),
\end{align}
where $M_{u},M_{d},M_{\text{lp}}$ are the quarks and charged leptons diagonal mass matrices and $M_D$ is the Dirac neutrino mass matrix defined after Eq.~(\ref{Eq:numassmtrxis}) and solved for it in Eq.~(\ref{Eq:ismd}). According to Eqs.~(\ref{Eq:chfrcp7}) to (\ref{Eq:yfylt}), we can write the charged Higgs boson couplings to fermions in terms of the physical fermion masses as follows:
\begin{align}
\label{Eq:chfrcp9}
C_{ij}&\simeq \frac{\sqrt{2}}{v c_{2\beta}}
\sum_{a=1}^{3} V^{*}_{ja} \big( V M_{d} V^\dag - s_{2\beta} M_{u}\big)^{*}_{ai},\\
\label{Eq:chfrcp10}
D_{ij}&\simeq \frac{\sqrt{2}}{v c_{2\beta}}
\sum_{a=1}^{3} V_{ja} \big( s_{2\beta}V M_{d} V^\dag - M_{u}\big)_{ia},\\
\label{Eq:chfrcp5}
\xi_{k\ell}&\simeq \frac{\sqrt{2}}{v c_{2\beta}}
\sum_{i=1}^3 U^*_{k,i+3} \big(s_{2\beta} M_{\text{lp}} - M_D\big)_{\ell{i}},\quad k=4,\ldots,9,\\
\label{Eq:chfrcp6}
\zeta_{k\ell}&\simeq \frac{\sqrt{2}}{v c_{2\beta}}
\sum_{i=1}^3 U_{ki} \big( M_{\text{lp}} - s_{2\beta} M_D\big)_{i\ell},\quad\quad~k=1,2,3,
\end{align}
where the conjugate ``*'' is omitted from the matrices when they are (taken) real. As noted after Eq.~(\ref{Eq:chfrcp4}), for $t_\beta\ll1~(t_\beta\gg1)$, the couplings $\xi_{k\ell}~(\zeta_{k\ell})$ are $M_D~(M_{\text{lp}})$-dominant and uncorrelated, and the family components are discriminant. Similarly, the above discussion applies for the charged Higgs boson couplings with quarks as well.

\noindent
Finally, we close this section by stating the scalar and pseudoscalar Higgs bosons sectors which were analyzed in detail with their couplings with charged leptons in~\cite{Ezzat:2021bzs}
\begin{align}
\Gamma^{h_i}_{\ell\ell}&=\frac{v}{\sqrt{2} m_\ell} \big(Z^H_{i1}\tilde{y}^L_{\ell\ell}+Z^H_{i2}y^L_{\ell\ell}\big),\\
\Gamma^{A}_{\ell\ell}&=\frac{v}{\sqrt{2} m_\ell} \big(Z^A_{31}\tilde{y}^L_{\ell\ell}-Z^A_{32}y^L_{\ell\ell}\big),
\end{align}
where $Z^H,Z^A$ are the scalar and pseudoscalar Higgs mixing matrices, respectively~\cite{86718,Ezzat:2021bzs}. More details about the LRIS Higgs and gauge sectors couplings and mixing and their parameters and spectra can be found in our previous work in Ref.~\cite{Ezzat:2021bzs}.

\section{\label{Sec:lrisamu}LRIS Contributions to Muon Anomalous Magnetic Moment}
\begin{figure}[t]
\centering
\vspace{-0.3in}
\includegraphics[trim={4.5cm 20.5cm 4cm 4cm},clip,scale=.9]{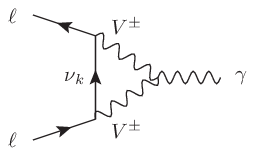}~\hspace{-2.9in}\quad
\includegraphics[trim={4.5cm 20.5cm 4cm 4cm},clip,scale=.9]{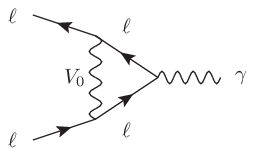}~\hspace{-2.9in}\quad
\includegraphics[trim={4.5cm 20.5cm 4cm 4cm},clip,scale=.9]{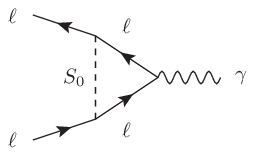}~\hspace{-2.9in}\quad
\includegraphics[trim={4.5cm 20.5cm 4cm 4cm},clip,scale=.9]{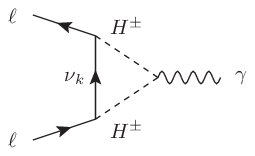}
\vspace{-0.7in}
\caption{\label{Fig:fyndgmgm2}LRIS one-loop Feynman diagrams contributions to lepton $g_\ell-2$ via massive neutrinos, $V^\pm=W,W',~V^0=Z,Z',~S^0=h_i,A$, and the charged Higgs boson $H^{\pm}$.}
\end{figure}

In this section we analyze new contributions from the LRIS to the muon anomalous magnetic moment, $a_\mu$, induced by the light and heavy $Z,W,Z',W'$ gauge bosons, as well as the neutral scalar and pseudoscalar and charged Higgs bosons $h,A,H^{\pm}$, as shown by their Feynman diagrams in Fig.~\ref{Fig:fyndgmgm2}. We will also consider the constraints on these contributions imposed by the experimental limits of the electron anomalous magnetic moment, $a_e$, and charged LFV processes, particularly, the $\mu\to{e}\gamma$ decay and the $\mu$-${e}$~conversion. In this case, we can write $\delta{a_\mu}= a_\mu^{\rm LRIS}$, where 
\begin{equation}
a_\mu^{\text{LRIS}}=a_\mu^W+a_\mu^{W'}+a_\mu^Z+a_\mu^{Z'}+a_\mu^{h}+a_\mu^{A}+a_\mu^{H^\pm}.
\end{equation}
The relevant amplitudes are, ignoring the $W-W'$ and $Z-Z'$ mixing, given by
\begin{align}
\label{Eq:amuwwpmn1}
a_{\ell}^{W}&=G^{\ell}_{\text{F}}~\sum_{k=1}^{9}|U_{k,|\ell|}|^2 \Big(\frac{10}{3}+\mathcal{F}_2(x^{\nu_{k}}_{W})\Big),\\
\label{Eq:amuwwpmn2}
a_\ell^{W'}&=G^{\ell}_{\text{F}}~\sum_{k=1}^{9}|U_{k,3+|\ell|}|^2 \Big(\frac{10}{3}+\mathcal{F}_2(x^{\nu_{k}}_{W'})\Big)\Big[\frac{1}{c_w} x^{W}_{W'}\Big],\\
\label{Eq:amuwwpmn3}
a_\ell^{Z}&=G^{\ell}_\text{F}~\Big(c_{4w}-5\Big)\big(\frac{1}{3}\big),\\
\label{Eq:amuwwpmn4}
a_\ell^{Z'}&=G^{\ell}_\text{F}~\Big(c_{4w'}-12 c_{2w'}-5\Big)\Big[\frac{t_w^2}{48s_{2w'}^2} x^{W}_{Z'}\Big],\\
\label{Eq:amuwwpmn6}
a_\ell^{h}&=G^{\ell}_\text{F}~\sum_{i=1}^3x^{\ell}_{h_i}(\Gamma^{h_i}_{\ell\ell})^2\Big(\frac{7}{6}+\log(x^{\ell}_{h_i})\Big),\\
\label{Eq:amuwwpmn7}
a_\ell^{A}&=G^{\ell}_\text{F}~\big(\frac{-1}{2}\big)x^{\ell}_{A}(\Gamma^{A}_{\ell\ell})^2\Big(\frac{7}{6}+\log(x^{\ell}_{A})\Big),
\end{align}
where the lepton family order $|\ell|=1,2$ for $\ell=e,\mu$. The dimensionless coupling $G^{\ell}_{\text{F}}=\frac{G_{\text{F}}m_\ell^2}{8\sqrt{2}\pi^2}$ and the mass ratio parameters $x^{a}_{b}=\frac{m_a^2}{m_b^2},~a=\nu_k,W,\ell,~b=W,W',Z',h_i,A,H^\pm$. The neutral gauge bosons mixing angles $\theta_{w'}$ and the Weinberg angle $\theta_w$ are $s_{w'} =\frac{g_Y}{g_R},~s_w =\frac{e}{g_L}$, where $g_Y$ is the hypercharge coupling. The $Z-Z'$ mixing angle $\theta_{w'}$ is constrained by $t_{w'}\lsim10^{-4}$~\cite{Pankov:2019yzr,Osland:2022ryb}. Also, the $W'$ mass is given by $m_{W'}=g_L\sqrt{v_R^2+v^2}/2\gsim\mathcal{O}(4~\text{TeV})$~\cite{Pankov:2019yzr,Osland:2022ryb}.

The analytical expressions of Eqs.~(\ref{Eq:amuwwpmn1})$-$(\ref{Eq:amuwwpmn7}) show that numerical values of the BSM contributions to $a_\mu$ mediated by $W',Z',h_i,A$ are negligible due to the suppression of their masses ratios $x^W_{W',Z'}$ and $x^\ell_{h_i,A}$, and thus we exclude their minor contributions. Also, the second summation term $\sum_{k=1}^{9}|U_{k,|\ell|}|^2~\mathcal{F}_2(x^{\nu_{k}}_{W})\simeq\sum_{k=4}^{9}|U_{k,|\ell|}|^2~\mathcal{F}_2(x^{\nu_{k}}_{W})$ of Eq.~(\ref{Eq:amuwwpmn1}) which represents the $W$-RHN loops contributions of Fig.~\ref{Fig:fyndgmgm2} is typically $\sim\mathcal{O}(10^{-2})$; only $\lsim0.4\%$ of the first term ($\frac{10}{3}$), and thus suppressed. This can be generally understood via the GIM cancellation mechanism~\cite{PhysRevD.2.1285} due to the unitarity of the full $9\times9$ neutrino mixing matrix $U$ within the nonunitarity limits of the $3\times3$ $U_{\text{PMNS}}$ light neutrino mixing matrix~\cite{Antusch:2006vwa,Malinsky:2009df,Ibarra:2011xn,Dev:2009aw,Abdallah:2011ew,Esteban:2020cvm}. In light of this, it can be generally concluded that any minimal BSM extension of the SM with RHN and with any adopted seesaw mechanism can not account for the measured $a_\mu$ anomaly and extra degrees of freedom are needed for this~\cite{Abdallah:2011ew}. The LRIS with its extra degrees of freedom is a good candidate for such class of BSM models.

\begin{figure}[t]
\centering
\includegraphics[scale=.9]{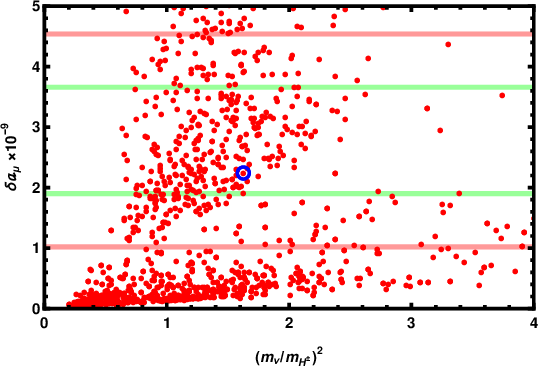}~\quad~\includegraphics[scale=.9]{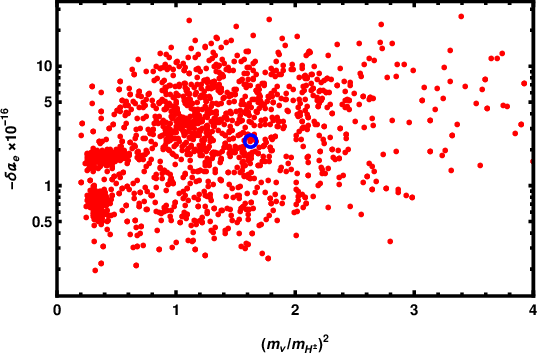}
\caption{\label{Fig:delamuaemnumch}Left/right: the muon/electron magnetic moment anomalies $\delta{a_\mu},-\delta{a_e}$ versus of the second heaviest neutrino and charged Higgs boson masses ratio parameter $x^{\nu_5}_{H^\pm}=m_{\nu_5}^2/m_{H^\pm}^2$. The $1\sigma$ and $2\sigma$ standard errors of measurements of $\delta{a_\mu}$ are included in green and red borders, respectively. The BP of Table~\ref{Tab:BPs1} is encircled.}
\end{figure}

\noindent
Finally, the charged Higgs boson $H^\pm$ contribution to $a_\mu$ is given by 
\begin{align}
\label{Eq:amuchmn}
a_\ell^{H^\pm}&=
G^{\ell}_{\text{F}}~\Gamma^{H^\pm}_{\gamma}~
\sum_{k=1}^{9}\Big(
|{\zeta'}_{k\ell}|^2~\mathcal{F}_2(x^{\nu_k}_{H^\pm})
+2\text{Re}[{\zeta'}_{k\ell} {\xi'}_{k\ell}^{*}]~\mathcal{F}_1(x^{\nu_k}_{H^\pm})
\Big),
\end{align}
where the charged Higgs boson interaction couplings with leptons $\xi_{k\ell},\zeta_{k\ell}$ appear in Eq.~(\ref{Eq:chfrcp21}), and ${\zeta'}_{k\ell}=\frac{v}{m_{\nu_k}}\zeta_{k\ell}$ and ${\xi'}_{k\ell}=\frac{v}{m_\ell}\xi_{k\ell}$.
The charged Higgs boson interaction coupling with photons is
\begin{align}
\label{Eq:xmxnzg3}
\Gamma^{H^\pm}_{\gamma}
&=\frac{1}{6e}
\Big(g_L U^{0}_{21}+g_R U^{0}_{31}+\big(g_{BL} U^{0}_{11}-g_L U^{0}_{21}\big)\big(Z^{H^\pm}_{33}\big)^2\Big)
\simeq
\frac{1}{6e}
\Big(g_L U^{0}_{21}+g_R U^{0}_{31}\Big),
\end{align}
where the last approximation is for $v_R\gg v$ where $Z^{H^\pm}_{33}\ll1$, as noted before Eq.~(\ref{Eq:chcomb2}). The matrix $U^{0}$ is the neutral gauge bosons mixing matrix and $g_R$ is the $SU(2)_R$ coupling~\cite{Ezzat:2021bzs}. The loop functions $\mathcal{F}_k~(k=1,2)$ in Eqs.~(\ref{Eq:amuwwpmn1}),~(\ref{Eq:amuwwpmn2})~and~(\ref{Eq:amuchmn}) are given by 
\begin{align}
\label{Eq:amulpfn1}
\mathcal{F}_k(y)&=\frac{y \mathcal{P}_{k}(y)}{(y-1)^{k+1}}-\frac{6y^{k+1}\log(y)}{(y-1)^{k+2}},\quad k=1,2,\\
\label{Eq:amulpfn2}
\mathcal{P}_{1}(y)&=3y+3,\\
\label{Eq:amulpfn3}
\mathcal{P}_{2}(y)&=2y^2+5y-1.
\end{align}
It is understood that for $y\to1$, the values of the loop functions $\mathcal{F}_k~(k=1,2)$ are given by their limits and $\mathcal{F}_1(1)=1$ and $\mathcal{F}_2(1)=\frac{1}{2}$. This happens when some heavy neutrinos are degenerate in mass with the charged Higgs boson as in Fig.~\ref{Fig:delamuaemnumch}. Asymptotically, the ratio $\mathcal{F}_2(x)/\mathcal{F}_1(x)$ is increasing and bounded below and above, and $\mathcal{F}_2(x),\mathcal{F}_1(x)\to0$ for $x\ll1$ such that $\mathcal{F}_2(x)/\mathcal{F}_1(x)\to1/3$ for $x\ll1$, and $\mathcal{F}_2(x)/\mathcal{F}_1(x)\to2/3$ for $x\gg1$. Accordingly, the two loop functions $\mathcal{F}_1$ and $\mathcal{F}_2$ remain of the same order for all possible values of the argument $x$.
Typically, the coupling $\Gamma^{H^\pm}_{\gamma}\sim0.076$. Also, the first contribution term of Eq.~(\ref{Eq:amuchmn}) $\sum_{k=1}^{9}|{\zeta'}_{k\ell}|^2~\mathcal{F}_2(x^{\nu_k}_{H^\pm})\simeq\sum_{k=1}^{3}|{\zeta'}_{k\ell}|^2~\mathcal{F}_2(x^{\nu_k}_{H^\pm})$ is $\sim\mathcal{O}(10^{-13})$. For light neutrinos, this term is suppressed by the loop function $\mathcal{F}_2(x^{\nu_\ell}_{H^\pm})$, while for heavy neutrinos, it is suppressed by their squared masses in the denominators of coefficients $|{\zeta'}_{k\ell}|^2~(k=4,\ldots,9)$. Indeed, the first term represents only $0.02\%$ of the second term $\sum_{k=1}^{9}2\text{Re}[{\zeta'}_{k\ell} {\xi'}_{k\ell}^{*}]~\mathcal{F}_1(x^{\nu_k}_{H^\pm})$$\simeq$$\sum_{k=4}^{9}2\text{Re}[{\zeta'}_{k\ell} {\xi'}_{k\ell}^{*}]~\mathcal{F}_1(x^{\nu_k}_{H^\pm})$, which is $\mathcal{O}(10^{-9})$, where this time, the second term is enhanced due to the charged lepton masses in the denominators of the coefficients ${\zeta'}_{k\ell} {\xi'}_{k\ell}^{*}$. Thus, the charged Higgs boson contribution to the $a_\mu$ anomaly Eq.~(\ref{Eq:amuchmn}) can be approximated to
\begin{align}
\label{Eq:amuchmn1}
a_\ell^{H^\pm}&\simeq
2G^{\ell}_{\text{F}}~\Gamma^{H^\pm}_{\gamma}~
\sum_{k=4}^{9} \text{Re}[{\zeta'}_{k\ell} {\xi'}_{k\ell}^{*}]~\mathcal{F}_1(x^{\nu_k}_{H^\pm})
%
\lsim
\frac{3\Gamma^{H^\pm}_{\gamma}}{8\pi^2}~m_\ell
\sum_{k=4}^{9} \frac{{\zeta}_{k\ell} {\xi}_{k\ell}}{m_{\nu_k}},
\end{align}
where $\frac{3\Gamma^{H^\pm}_{\gamma}}{8\pi^2}\sim3\times10^{-3}$ and the loop function $\mathcal{F}_1$ is increasing and bounded above such that $\mathcal{F}_1(x)\to3$ for $x\gg1$, and the complex notations ``$\text{Re},~*$'' are omitted as the couplings are (taken) real.

In the rest of this section, we analyze the parameter space of the LRIS for numerical scan for benchmark points (BPs). In the SM, all particles acquire their masses via the VEV of only one degree of freedom, the Higgs field, and each particle mass depends only on one parameter coupling, its coupling with the Higgs field. This feature almost fixes the SM parameters values, except maybe due to some measurements uncertainties. So, in the SM, couplings are fixed at the EW scale by particles masses. Conversely, in LRIS, there are many sources of VEVs and couplings for particles' masses. So, in LRIS, VEVs, the Yukawa couplings and scalar potential parameters are in general free parameters (see Ref.~\cite{Ezzat:2021bzs}), and they can be varied while fermions and scalar masses are kept fixed. Also, in LRIS, the gauge couplings are constrained by the gauge bosons masses at the EW scale and by their renormalization group equations (RGEs) evolution up to GUT scale, especially when we fix $v_R\sim\mathcal{O}(10~\text{TeV})$ for the extra $W',Z'$ experimental mixings and masses limits as discussed after Eq.~(\ref{Eq:amuwwpmn7}).

The neutrinos masses Eqs.~(\ref{Eq:nulmass})~and~(\ref{Eq:nuhmass}) and their mixing matrix $U$ after Eq.~(\ref{Eq:numassmtrxis}) are given in terms of $M_D,M_R$, and $\mu^s$ (or equivalently $y^L,\tilde{y}^L,y^s,\mu^s,t_\beta$, and $v_R$). In our numerical analysis, we fix $v_R\sim\mathcal{O}(10~\text{TeV})$. Also, we adopted the normal hierarchy of light neutrino masses $m_{\nu_\ell}$, as in Table~\ref{Tab:BPs1}. The chosen light neutrino masses values satisfy 
$\Delta{m}_{21}^2=7.224\times10^{-5}~\text{eV}^2$ and $\Delta{m}_{31}^2=2.500\times10^{-3}~\text{eV}^2$.  They agree with the $1\sigma$ ranges of the observed solar and atmospheric mass splittings values $\Delta{m}_{\text{sol}}^2=7.420_{-0.200}^{+0.210}\times10^{-5}~\text{eV}^2$, and $\Delta{m}_{\text{atm}}^2=2.517_{-0.028}^{+0.026}\times10^{-3}~\text{eV}^2$~\cite{Esteban:2020cvm}.
But, to satisfy the $a_\mu$ anomaly in the inverted hierarchy scenario of neutrino masses within the imposed LFV constraints, the nonunitarity limits of the $U_{\text{PMNS}}$ matrix should be violated. In this case, one has to set relatively large ($\sim\mathcal{O}(10^{-10})$) and thus almost degenerate light neutrino masses to satisfy the inverted hierarchy scenario observed limits $\Delta{m}_{21}^2=\Delta{m}_{\text{sol}}^2=7.420_{-0.200}^{+0.210}\times10^{-5}~\text{eV}^2$, and $\Delta{m}_{32}^2=\Delta{m}_{\text{atm}}^2=-2.498_{-0.028}^{+0.028}\times10^{-3}~\text{eV}^2$~\cite{Esteban:2020cvm}. Also, $\mu^s$ is enlarged for the $a_\mu$ anomaly and other relevant LFV constraints, and it mixes with other entries in the neutrino mass matrix~(\ref{Eq:numassmtrxis}), thus violating the nonunitarity limits of $U_{\text{PMNS}}$~\cite{Esteban:2020cvm}.
We chose to express $M_D$ (and hence, in correlation, $y^L,\tilde{y}^L$) in terms of $y^s,\mu^s$ and $t_\beta$ as in Eqs.~(\ref{Eq:ismd}),~(\ref{Eq:yfyl})~and~(\ref{Eq:yfylt}). Accordingly, substituting $M_D$ and $y^s$ in the heavy neutrino masses Eq.~(\ref{Eq:nuhmass}) determine them. So, in our numerical analysis below, we fix the normal hierarchy light neutrino masses and the entries of the $U_{\text{PMNS}}$ matrix and scan over $y^s$ and $\mu^s$ as in Eq.~(\ref{Eq:prmrngs}) for the neutrino sector~\cite{Esteban:2020cvm}. It is also worth mentioning here that the consistency of all numerical calculations of Table~\ref{Tab:BPs1} is verified. For example, the $M_D$ matrix, when calculated from its main definition after Eq.~(\ref{Eq:numassmtrxis}) is found consistent with its values from Eq.~(\ref{Eq:ismd}). Also, the neutrino masses in Table~\ref{Tab:BPs1} are consistent with Table~\ref{Tab:BPs2} and Eqs.~(\ref{Eq:nulmass}),~(\ref{Eq:nuhmass})~and~(\ref{Eq:upmns}). Finally, the orthogonal matrix $\mathcal{R}$ of Eq.~(\ref{Eq:ismd}) was fixed such that its nonvanishing components are $\mathcal{R}_{13}=\mathcal{R}_{21}=\mathcal{R}_{32}=1$ as given in Table~\ref{Tab:BPs2}.

Also, the charged Higgs boson mass Eq.~(\ref{Eq:chm}) is varied versus $t_\beta$ and the scalar potential paremeter $\alpha$, and the charged Higgs mixing $Z^{H^\pm}$ Eq.~(\ref{Eq:chhgsmix}) is given in terms of $t_\beta$ and $v_R$. For the charged Higgs boson mass and mixing, we scan over $t_\beta$ and $\alpha$ as in Eq.~(\ref{Eq:prmrngs}).
As detailed above, after Eqs.~(\ref{Eq:nuhmass}),~(\ref{Eq:chm})~and~(\ref{Eq:chfrcp4}), in our numerical analysis of the neutrino and charged Higgs sectors, we scanned over the following independent parameters' ranges [with $v_R\sim\mathcal{O}(10~\text{TeV})$]
\begin{equation}\label{Eq:prmrngs}
\alpha\sim[0.005,0.050],\quad t_{\beta}\sim[0.01,0.99],\quad (y^s)_{ij}\sim[0.01,0.50]\delta_{ij},\quad (\mu^s)_{ij}\sim[10^{-9},10^{-5}]\delta_{ij}~\text{GeV}.
\end{equation}
We checked that all of our BPs are validated to satisfy the usual \higgsbounds~and \higgssignals~limits confronted with the latest LEP, Tevatron, and LHC data~\cite{Bechtle:2013wla,Bechtle:2013xfa}. They provide important tests for compatibility of any BSM model. In our analysis, the LRIS model was first built in the \sarah~package, then it was passed to \spheno~for numerical spectrum calculations~\cite{Porod:2011nf,Staub:2013tta}. Specifically, we present one of our BPs in Table~\ref{Tab:BPs1} with the corresponding observables in Table~\ref{Tab:BPs3} and other parameters in Table~\ref{Tab:BPs2} and Eq.~(\ref{Eq:upmns}).

\begin{figure}[t]
\centering
\includegraphics[scale=.9]{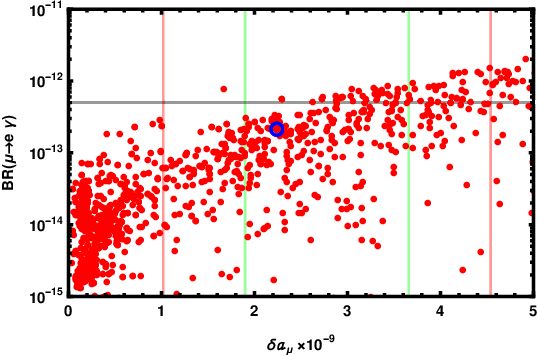}
\caption{\label{Fig:mu2egmamnumchdlamu}The branching ratio $\text{BR}(\mu\to{e}\gamma)$ versus the muon anomalous magnetic moment deviation $\delta{a_\mu}$ in LRIS. The green (red) borders are the $1\sigma~(2\sigma)$ standard error of measurements of $\delta{a_\mu}$, and the gray horizontal line is the upper bound on $\text{BR}(\mu\to{e}\gamma)$. The BP of Table~\ref{Tab:BPs1} is encircled.}
\end{figure}

The left (right) panel of Fig.~\ref{Fig:delamuaemnumch} depicts the muon (electron) $g_{\mu(e)}-2$ anomalies $\delta{a_{\mu(e)}}$ in LRIS, as given in Eqs.~(\ref{Eq:amuwwpmn1})~and~(\ref{Eq:amuchmn}), resulting from the BSM contributions of the $W$-RHN loops and the charged Higgs boson contribution. We choose, without any loss of generality or independence, to show the distribution of $\delta{a_{\mu(e)}}$ versus the mass ratio $x^{\nu_5}_{H^\pm}=m_{\nu_5}^2/m_{H^\pm}^2$ for its moderate and variable values, as it is clear from the masses values BP of Table~\ref{Tab:BPs1}, but any other of the independent or dependent parameters or any one of the heavy neutrinos ratios $x^{\nu_j}_{H^\pm}~(j=4,\ldots,9)$ would equally work for the same set of data. The green (red) borders indicate the $1\sigma~(2\sigma)$ level of accuracy around the average $\delta{a_\mu}$ as in Eq.~(\ref{Eq:g2mu}). The electron anomaly $\delta{a_e}$ is guaranteed to be within the allowed experimental uncertainty limits $|\delta{a_e}|\lsim(10^{-15}-10^{-13})$~\cite{Giudice:2012ms,NA64:2021xzo}. So, all BPs used in Figs.~\ref{Fig:delamuaemnumch}~and~\ref{Fig:mu2egmamnumchdlamu} satisfy the electron $a_e$ anomaly limits.
Furthermore, Fig.~\ref{Fig:mu2egmamnumchdlamu} shows that the LVF $\text{BR}(\mu\to{e}\gamma)$ in Eq.~(\ref{Eq:BRmu2egma}) satisfies the experimental bounds for the same set of parameters values as in Fig.~\ref{Fig:delamuaemnumch}.
\begin{table}[t]
\begin{center}
\begin{tabular}{cccc|cccc|ccccccccc}
\hline
\hline
\textbf{Par} & $\alpha$ & $t_{\beta}$ & $v_R$ & $Z^{H^\pm}_{31}$ & $Z^{H^\pm}_{32}$ & $Z^{H^\pm}_{33}$ & $m_{H^\pm}$ & $m_{\nu_1}$ & $m_{\nu_2}$ & $m_{\nu_3}$ & $m_{\nu_4}$ & $m_{\nu_5}$ & $m_{\nu_6}$ & $m_{\nu_7}$ & $m_{\nu_8}$ & $m_{\nu_9}$ \\\hline
\textbf{BP} & $0.0058$ & 0.1 & 10000 & $-0.099$ & $-0.994$ & $0.024$ & 545 & $1.0\times 10^{-13}$ & $8.5\times 10^{-12}$ & $5.0\times 10^{-11}$ & 108 & 695 & 1449 & 108 & 695 & 1449 \\
\hline
\hline
\end{tabular}
\caption{\label{Tab:BPs1}BP and corresponding charged Higgs boson and neutrino mass spectrum in GeVs.}
\end{center}
\end{table}
\begin{table}[t]
\begin{center}
\begin{tabular}{cccc|cccc|c}
\hline
\hline
\textbf{Matrix} & $\mathcal{R}$ & $y^s$ & $\mu^s$ & $y^L$ & $\tilde{y}^L$ & $y^Q$ & $\tilde{y}^Q$ & $U_{\text{PMNS}}$ \\ 
\hline
$1,1$ & 0 & $1.53\times10^{-2}$ & $1.01\times10^{-5}$ & $-2.83\times10^{-5}$ & $~~3.13\times10^{-4}$ & $~~6.33\times10^{-5}$ & $-3.44\times10^{-4}$ & $~~0.8251$ \\ 
\hline
$1,2$ & 0 & 0 & 0 & $-6.86\times10^{-3} $ & $~~6.86\times10^{-2} $ & $-1.49\times10^{-4}$ & $~~1.48\times10^{-3}$& $~~0.5449$ \\ 
\hline
$1,3$ &1 & 0 & 0 & $-9.42\times10^{-5}$ & $~~9.42\times10^{-4}$ & $-3.53\times10^{-4}$ & $~~3.53\times10^{-3}$& $~~0.1490$ \\ 
\hline
$2,1$ &1 & 0 & 0 & $-2.75\times10^{-5}$ & $~~2.75\times10^{-4}$ & $~~1.38\times10^{-4}$ & $-1.38\times10^{-3}$& $-0.4554$ \\ 
\hline
$2,2$ & 0 & $9.76\times10^{-2}$ &$3.82\times10^{-9}$ & $-3.39\times10^{-2} $ & $~~3.46\times10^{-1}$ & $~~2.27\times10^{-5}$ &$~~5.26\times10^{-3}$& $~~0.4795$ \\ 
\hline
$2,3$ & 0 & 0 & 0 & $~~5.20\times10^{-5} $ & $-5.20\times10^{-4}$ & $-4.19\times10^{-3}$ & $~~4.19\times10^{-2}$& $~~0.7513$ \\ 
\hline
$3,1$ & 0 & 0 & 0 & $~~3.93\times10^{-4}$ & $-3.93\times10^{-3}$& $-6.08\times10^{-4}$ & $~~6.08\times10^{-3}$& $~~0.3343$ \\ 
\hline
$3,2$ &1 & 0 & 0 & $-2.96\times10^{-2}$ & $~~2.96\times10^{-1}$ & $~~4.16\times10^{-3}$ & $-4.16\times10^{-2}$& $-0.6836$ \\ 
\hline
$3,3$ & 0 & $2.05\times10^{-1}$ &$5.49\times10^{-6}$ & $~~1.03\times10^{-2}$ & $-6.54\times10^{-4}$ & $-7.66\times10^{-2}$ & $~~9.99\times10^{-1}$& $~~0.6427$ \\ 
\hline
\hline
\end{tabular}
\caption{\label{Tab:BPs2}Yukawa and IS matrices BP of Table~\ref{Tab:BPs1}.}
\end{center}
\end{table}
\begin{table}[t]
\begin{center}
\begin{tabular}{cccccccccc} 
\hline
\hline
\textbf{Quantity}	& $\delta{a_\mu}$		& $-\delta{a_e}$		& $\text{BR}(\mu\to{e}\gamma)$	& $R_{\mu\to{e}}^{\text{Al}}$	& $R_{\mu\to{e}}^{\text{Ti}}$	& $R_{\mu\to{e}}^{\text{Au}}$ \\\hline
\textbf{Value}		& $2.24\times{10^{-9}}$ & $2.30\times{10^{-16}}$& $2.10\times{10^{-13}}$		& $4.10\times{10^{-51}}$		& $3.80\times{10^{-50}}$		& $4.10\times{10^{-49}}$ \\
\hline
\hline
\end{tabular}
\caption{\label{Tab:BPs3}Results of muon and electron $g_{\mu(e)}-2$ and LFV processes $\text{BR}(\mu\to{e}\gamma)$ and $\mu$-${e}$~conversion rates of the BP given in Table~\ref{Tab:BPs1}.}
\end{center}
\end{table}

\noindent
The $9\times9$ neutrino mixing matrix $U$of the BP of Table~\ref{Tab:BPs1}, rounded to $\mathcal{O}(10^{-4})$, with the $U_{\text{PMNS}}$ matrix~\cite{Esteban:2020cvm},  is
\begin{equation}\label{Eq:upmns}
U
=\begin{pmatrix}
U_{3\times3}&U_{3\times6}\\
U_{6\times3}&U_{6\times6}
\end{pmatrix}^T
=\left(\begin{array}{ccc|cccccc}
-0.8243	&	~~0.4535&	-0.3389	&	0		&	0		&	0		&	0		&	~~0.0000&	-0.0001	\\
~~0.5465&	~~0.4812&	-0.6853	&	0		&	0		&	0		&	0.0009	&	~~0.0002&	0		\\
-0.1468	&	-0.7453	&	-0.6403	&	0		&	0		&	0		&	0		&	-0.1137	&	0		\\
\hline
-0.0004	&	-0.0003	&	~~0.0004&	-0.7071	&	0		&	0		&	0.7071	&	0		&	0		\\
-0.0120	&	-0.0604	&	-0.0517	&	0		&	-0.7071	&	0		&	0		&	~~0.7025&	0		\\
~~0.0001&	~~0.0000&	~~0.0003&	0		&	0		&	0.7071	&	0		&	0		&	-0.7071	\\
-0.0004	&	-0.0003	&	~~0.0004&	~~0.7071&	0		&	0		&	0.7071	&	0		&	0		\\
~~0.0120&	~~0.0604&	~~0.0517&	0		&	-0.7071	&	0		&	0		&	-0.7025	&	0		\\
-0.0001	&	~~0.0000&	~~0.0000&	0		&	0		&	0.7071	&	0		&	0		&	~~0.7071\\
\end{array}\right),
\end{equation}
where each block matrix is parametrized such that~\cite{Abdallah:2011ew,Dev:2009aw}
\begin{align}\label{Eq:upmns2}
U_{3\times6}&\simeq\big[\mathbf{0}_{3\times3}|F\big]_{3\times6}~U_{6\times6},\\
U_{3\times3}&\simeq\big(\mathbf{1}_{3\times3}-\frac{1}{2}F F^T\big)U_{\text{PMNS}},
\end{align}
where the nonunitarity limits of the $U_{\text{PMNS}}$ is encoded in $F=M_DM_R^{-1}$ and the $3\times6$ extended matrix $\mathbb{F}=\big[\mathbf{0}_{3\times3}|F\big]_{3\times6}$ is $\mathbb{F}_{ij}=0$ and $\mathbb{F}_{i,j+3}=F_{ij}$ for $i,j=1,2,3$. Finally, $U_{6\times6}$ is the matrix which diagonalizes the $\{\nu_R,S_2\}$ mass matrix, and we have omitted the standard global phase matrix which gives the true positive values of neutrino masses.

\section{\label{Sec:lrisLFV}Lepton Flavor Violation Constraints}

Now, we turn to the constraints on the charged Higgs boson contributions to LFV rare processes. The LRIS $W$-RHN and the charged Higgs boson contributions to the $\text{BR}(\mu\to{e}\gamma)$ and the $\mu$-${e}$~conversion rates $R_{\mu\to{e}}$ are in order. Experiments set upper bounds to these quantities, and the stringent experimental limits on these processes should be regarded as constraints on the charged Higgs boson contribution to $a_\mu$~\cite{Alonso:2012ji,Davidson:2018kud}. The LFV experiments set the upper limit $\text{BR}(\mu\to{e}\gamma)\lsim4.2\times10^{-13}$ with $90\%$ confidence level~\cite{MEG:2016leq}. In LRIS, the $W$-RHN and charged Higgs boson mediation for $\mu\to{e}\gamma$ leads to
{\small
\begin{align}\label{Eq:BRmu2egma}
\text{BR}(\mu\to{e}\gamma)_{\text{LRIS}}=
\frac{\alpha^3_w s_w^2}{256\pi^2} \frac{m_\mu}{\Gamma_\mu} (x^{\mu}_{W})^2
\sum_{k=1}^{9}
\Big|
U_{k,1}U_{k,2}^*~\mathcal{F}_2(x^{\nu_{k}}_{W})
+{\zeta'}_{k,e} {\zeta'}_{k,\mu}^*~\mathcal{F}_2(x^{\nu_k}_{H^\pm})
+\big({\zeta'}_{k,e} {\xi'}_{k,\mu}^*+{\xi'}_{k,e} {\zeta'}_{k,\mu}^*\big)~\mathcal{F}_1(x^{\nu_k}_{H^\pm})
\Big|^2.
\end{align}
}
As discussed after Eq.~(\ref{Eq:amuwwpmn7}), the first term $\sum_{k=1}^{9}U_{k,1}U_{k,2}^*~\mathcal{F}_2(x^{\nu_{k}}_{W})\simeq\sum_{k=1}^{3}{(U_{\text{PMNS}}^T)}_{k,1}{(U_{\text{PMNS}}^T)}_{k,2}~\mathcal{F}_2(x^{\nu_{k}}_{W})$ of Eq.~(\ref{Eq:BRmu2egma}) of the $W-\nu$ contribution is $\lsim\mathcal{O}(10^{-29})$, and thus negligible by the GIM cancellation mechanism~\cite{PhysRevD.2.1285}. The remaining $W$-RHN contribution in the first term $\sum_{k=4}^{9}U_{k,1}U_{k,2}^*~\mathcal{F}_2(x^{\nu_{k}}_{W})$ vanishes due to contributions from the first two rows of the $3\times6$ upper-right block matrix in Eq.~(\ref{Eq:upmns}) which gives $U_{k,1}U_{k,2}^*\approxeq0,~k=4,\ldots,9$. Accordingly, we only constrain the charged Higgs boson contribution. For this, as discussed in the paragraph before Eq.~(\ref{Eq:amuchmn1}), the second term
$\sum_{k=1}^{9}{\zeta'}_{k,e} {\zeta'}_{k,\mu}^*~\mathcal{F}_2(x^{\nu_k}_{H^\pm})\simeq
\sum_{k=1}^{3}{\zeta'}_{k,e} {\zeta'}_{k,\mu}^*~\mathcal{F}_2(x^{\nu_k}_{H^\pm})$ is $\sim\mathcal{O}(10^{-9})$,
and it is only about $0.004\%$ of the third term
$\sum_{k=1}^{9}({\zeta'}_{k,e} {\xi'}_{k,\mu}^*+{\xi'}_{k,e} {\zeta'}_{k,\mu}^*)~\mathcal{F}_1(x^{\nu_k}_{H^\pm})$
$\simeq
\sum_{k=4}^{9}({\zeta'}_{k,e} {\xi'}_{k,\mu}^*+{\xi'}_{k,e} {\zeta'}_{k,\mu}^*)~\mathcal{F}_1(x^{\nu_k}_{H^\pm})$, which is $\sim\mathcal{O}(10^{-5})$ and need to be constrained. We can approximate Eq.~(\ref{Eq:BRmu2egma}) as
\begin{align}\label{Eq:BRmu2egma1}
\text{BR}(\mu\to{e}\gamma)_{\text{LRIS}}&\simeq
\frac{\alpha^3_w s_w^2}{256\pi^2} \frac{m_\mu}{\Gamma_\mu} (x^{\mu}_{W})^2
\sum_{k=4}^{9}
\Big|
\big({\zeta'}_{k,e} {\xi'}_{k,\mu}^*+{\xi'}_{k,e} {\zeta'}_{k,\mu}^*\big)~\mathcal{F}_1(x^{\nu_k}_{H^\pm})\Big|^2\nonumber\\
&\lsim
\frac{9\alpha_{\text{em}}}{256\pi^4} \frac{m_\mu^5}{\Gamma_\mu} 
\sum_{k=4}^{9} \frac{1}{m_{\nu_k}^2}\Big(\frac{{\zeta}_{k,e} {\xi}_{k,\mu}}{m_\mu}+\frac{{\xi}_{k,e} {\zeta}_{k,\mu}}{m_e}\Big)^2,
\end{align}
where the factor $\frac{9\alpha_{\text{em}}}{256\pi^4} \frac{m_\mu^5}{\Gamma_\mu}\sim10^{8}$ and the loop function $\mathcal{F}_1(x)\lsim3$ for $x\gg1$, as noted after Eq.~(\ref{Eq:amuchmn1}), and again the complex notations are omitted as the couplings are (taken) real.

At the end, we check experimental limits on the charged Higgs boson contributions to the $\mu$-${e}$~conversion on a nucleus with atomic weight $A$. The charged Higgs contributes to the $\mu$-${e}$~conversion rate as follows~\cite{Alonso:2012ji,Davidson:2018kud}
\begin{equation}\label{Eq:mu2e}
R^{A}_{\mu\to{e}}=\frac{32G_\text{F}^2 m_\mu^5}{\Gamma^{A}_{\text{capt}}}
\Big[
\Big|\widetilde{C}^{pp}_{V,R} V^{(p)}_{A} + \widetilde{C}^{nn}_{V,R} V^{(n)}_{A} + \frac{1}{4} C_{D,L} D_{A}\Big|^2 
+\{L\leftrightarrow R\}
\Big],
\end{equation}
where $\Gamma^{A}_{\text{capt}}$ is the rate for the muon to transform to a neutrino by capture on the nucleus $(A)$. Some numerical values of $\Gamma^{A}_{\text{capt}}\sim\mathcal{O}(1-10)\times10^{6}~s^{-1}$, and the nucleus and nucleon $(n,p)$-dependent ``overlap integrals" $V^{(p)}_{A},V^{(n)}_{A},D_{A}\sim\mathcal{O}(10^{-2}-10^{-1})$ for the nuclei $A=\text{Al,~Ti,~Au}$ are given in Ref.~\cite{Kitano:2002mt}. Experiments make the upper bounds
$R_{\mu\to{e}}^{\text{Ti}} \leqslant 10^{-18},~
R_{\mu\to{e}}^{\text{Al}} \leqslant 10^{-16},~
R_{\mu\to{e}}^{\text{Au}} \leqslant 7\times 10^{-13}.$
In LRIS, the nucleon-dependent Wilson coefficients are given by
\begin{align}
\label{Eq:mu2e1}
C_{D,L}&=\frac{8G_{\text{F}}\alpha_{\text{em}}}{\pi{s_w^2}\sqrt{2}}
\sum_{k=1}^{9}\sum_{j=1}^{3}\sum_{q,q'=u,d,q\neq{q'}} {\big(U_{k,e}^{*}U_{k,\mu}|V_{q',q_j}|^2\big)~B_2(x^{\nu_k}_{W},x^{q_j}_{W})}, \\
\label{Eq:mu2e2}
\widetilde{C}^{pp}_{V,R} &=\frac{1}{8\pi^2 m_{H^{\pm}}^2} 
\sum_{k=1}^{9}\sum_{j=1}^{3}\sum_{q,q'=u,d,q\neq{q'}} {\big(\zeta_{k,e}\xi_{k,\mu}+\zeta_{k,\mu}\xi_{k,e}\big) 
\big(C_{q',q_j}^2+D_{q',q_j}^2\big)~B_2(x^{\nu_k}_{H^\pm},x^{q_j}_{H^\pm}) }, \\
\label{Eq:mu2e3}
\widetilde{C}^{nn}_{V,R} &=\frac{1}{4\pi^2 m_{H^{\pm}}^2} 
\sum_{k=1}^{9}\sum_{j=1}^{3}\sum_{q,q'=u,d,q\neq{q'}} {\big(\zeta_{k,e}\zeta_{k,\mu}+\xi_{k,e}\xi_{k,\mu}\big) 
\big(C_{q',q_j}D_{q',q_j}\big)~B_1(x^{\nu_k}_{H^\pm},x^{q_j}_{H^\pm})\sqrt{x^{\nu_k}_{H^\pm} x^{q_j}_{H^\pm}} },
\end{align}
where in LRIS the interchange $\{L\leftrightarrow R\}$ does not change the coefficients. The parameters $C_{ij},D_{ij}$, and $\zeta_{k\ell},\xi_{k\ell}$ are the charged Higgs boson interaction couplings with quarks and leptons appearing in Eqs.~(\ref{Eq:chfrcp11})~and~(\ref{Eq:chfrcp21}), respectively, and the loop functions are
\begin{align}
\label{Eq:rklpfn2}
J_k(x)&=\frac{1}{1-x}+\frac{x^k\log(x)}{(1-x)^2},\\
\label{Eq:rklpfn}
B_k(x,y)&=\frac{J_k(x)-J_k(y)}{x-y},\quad k=1,2.
\end{align}
Asymptotically, $B_k(x,y)\to0$ as $x\gg1$ and $y\ll1$. So the $W-\nu$ contribution Eq.~(\ref{Eq:mu2e1}) is clearly suppressed by the GIM cancellation mechanism~\cite{PhysRevD.2.1285}. The factor $\frac{32G_\text{F}^2 m_\mu^5}{\Gamma^{A}_{\text{capt}}}\sim\mathcal{O}(10^{-21}-10^{-20})$ in Eq.~(\ref{Eq:mu2e}), and all BPs are tested and found to satisfy the $\mu$-${e}$~conversion experimental limits mentioned after Eq.~(\ref{Eq:mu2e}), and all of them are of order of the BP results in Table~\ref{Tab:BPs3}.

\section{\label{Sec:conclsn}Conclusion}
We have analyzed the muon anomalous magnetic moment $a_\mu$ in a minimal left-right symmetric model with neutrino masses inverse seesaw mechanism. We found that a reasonable region of the parameter space of the model is consistent with the observed muon $g-2$ anomaly. We emphasized that, in this type of models, only the $H^{\pm}$ loop explains $a_\mu$ significantly, in agreement with the $\text{BR}(\mu\to{e}\gamma)$, $\mu$-${e}$~conversion and the electron $g_e-2$ anomaly measured limits.

\section*{acknowledgements}
The work of M. A. is partially supported by Science, Technology $\&$ Innovation Funding Authority (STDF) under Grant No. 33495, and the work of K. E. and S. K. is partially supported by STDF under Grant No. 37272.

\bibliographystyle{apsrev}\bibliography{Bib}
\end{document}